\documentclass[aps,preprint,preprintnumbers,nofootinbib,showpacs]{revtex4}
\usepackage{graphicx,color,amsmath}
\begin{document}
\title{Radiative Baryonic $B$ Decays}
\author{C. Q. Geng and Y. K. Hsiao}
\affiliation{Department of Physics, National Tsing Hua University,
Hsinchu, Taiwan 300}
\date{\today}
\begin{abstract}
We study the structure-dependent contributions to the radiative
baryonic $B$ decays of $B \to {\bf B}{\bf \bar B'}\gamma$ in the
standard model. We show that the decay branching ratios of $Br(B \to
{\bf B}{\bf \bar B'}\gamma)$ are $O(10^{-7})$, which are larger
than the estimated values of $O(10^{-9})$ induced from inner
bremsstrahlung effects of the corresponding two-body modes. In
particular, we find that $Br(B^- \to \Lambda \bar p \gamma)$
is around $1 \times 10^{-6}$, which is close to the pole model estimation
but smaller than the experimental measurement from BELLE.
\end{abstract}
\newpage
\preprint{}
\maketitle

The radiative baryonic $B$ decays of $B \to {\bf B}{\bf \bar
B'}\gamma$ are of interest since they are three-body decays with
two spin-1/2 baryons (${\bf B}$ and ${\bf B'}$)
and one spin-1 photon in the final states.
The rich spin structures allow us to explore various interesting
observables such as triple momentum correlations to
 investigate CP or T violation \cite{threshold,TV}.
 Moreover, since these radiative decays could dominantly arise from
the short-distance electromagnetic penguin transition of $b \to
s\gamma$ \cite{reviewbtosg} which has been utilized to place
significant constraints on physics beyond the Standard Model (SM)
\cite{NB1,NB2}, they then appear to be the potentially applicable
probes to new physics.

There are two sources to produce radiative baryonic $B$ decays.
One is the inner bremsstrahlung (IB) effect, in which
the
radiative baryonic $B$ decays of $B \to{\bf B}{\bf \bar B'}\gamma$
are from their two-body decay counterparts of  $B \to{\bf B}{\bf \bar
B'}$ via the supplementary emitting photon attaching to one of the
final baryonic states. Clearly, the radiative decay rates due to the
IB contributions  are suppressed by $\alpha_{em}$ comparing
with their counterparts.
 According to the existing upper bounds of $B\to {\bf B}{\bf \bar{B'}}$, given by
 \cite{BtoPP,BtoBB,BtoLambdaP}
\begin{eqnarray}
 Br(\bar{B}^0 \to p\bar{p})
&<&2.7\times10^{-7}\;\text{(BABAR)}\;,\nonumber\\
Br(\bar{B}^0 \to \Lambda\bar{\Lambda})
&<&7.9\times 10^{-7}\;\text{(BELLE)},\nonumber\\
Br(B^- \to \Lambda\bar{p})
&<&4.6\times 10^{-7}\;\text{(BELLE)}\;,
\end{eqnarray}
one finds that
 \begin{eqnarray}
Br(B \to{\bf B}{\bf \bar B'}\gamma)_{IB}\leq O(10^{-9}).
\end{eqnarray}
Unfortunately, the above branching ratios are  far from the
present accessibility at the B factories of BABAR and BELLE.
However, the other source, which is the structure-dependent (SD),
is expected to enhance the decays of $Br(B \to{\bf B}{\bf \bar
B'}\gamma)$, such as $B\to\Lambda \bar{p}\gamma$ arising from $b
\to s \gamma$ \cite{threshold,HY,Hou1}. With the large branching
ratio of $b \to s\gamma$ \cite{btosgpdg,btosg} in the range of
$10^{-4}$ we expect that $Br(B^- \to {\bf B}{\bf \bar B'}\gamma)$
could be as large as $Br(B^- \to {\bf B}{\bf \bar B'})$. In this
report, we shall concentrate on the SD contributions to $Br(B
\to{\bf B}{\bf \bar B'}\gamma)$.

To start our study, we must tackle the cumbersome transition matrix elements in
$B \to {\bf B}{\bf\bar B'}$.
As more and more experimental data on three-body decays
\cite{pppi,threebody1,threebody2} in recent years, the theoretical progresses
 are improved to resolve the transition matrix element
problems. One interesting approach is to use the pole model
\cite{HY3,transitionHY} through the intermediated particles and
another one is to rely on the QCD counting rules
\cite{QCD1,QCD,transition} by relating the transition matrix elements with
three form factors and fitting with experimental data. In Ref.
\cite{HY}, Cheng and Yang have worked out the radiative baryonic B
decays based on the pole model. In this paper, we handle the
transition matrix elements according to the QCD counting rules.
\begin{figure}[t!]
\centering
\includegraphics[width=4in]{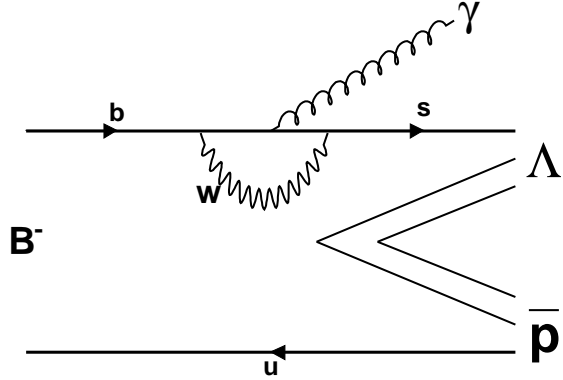}
\caption{\label{penguin} Diagram for $B^- \to \Lambda\bar p\gamma$}
\end{figure}

We begin with  the decay of $B^- \to \Lambda \bar p \gamma$. As
depicted in Fig. \ref{penguin}, in the SM the relevant
Hamiltonian due to the SD contribution for  $B^- \to \Lambda \bar
p\gamma $ is
\begin{eqnarray}
{\cal H}_{SD}=-\frac{G_F}{\sqrt 2}V_{tb}V_{ts}^*c^{eff}_7 O_7,
\end{eqnarray}
with the tensor operator
\begin{eqnarray}
O_7=\frac{e}{8\pi^2}m_b\bar
s\sigma_{\mu\nu}F^{\mu\nu}(1+\gamma_5)b,
\end{eqnarray}
where $V_{tb}V_{ts}^*$ and $c^{eff}_7$  are the CKM matrix
elements and Wilson coefficient, respectively, and  the decay
amplitude is found to be
\begin{eqnarray}\label{A1}
A(B^- \to \Lambda\bar p \gamma)&=& \frac{G_F}{\sqrt
2}V^*_{ts}V_{tb}\frac{e}{8\pi^2}2c^{eff}_7\nonumber\\
&& \bigg\{m_b^2\varepsilon^{\mu}\langle \Lambda\bar p|\bar s
\gamma_{\mu}(1-\gamma_5)b|B^-\rangle -2m_bp_B\cdot
\varepsilon\langle \Lambda\bar p|\bar
s(1+\gamma_5)b|B^-\rangle\bigg\},
\end{eqnarray}
where we have used the condition $m_b \gg m_s$ such that the terms
relating to $m_s$ are neglected. We note that Eq. (\ref{A1}) is
still gauge invariant.

In order to solve the encountered transition matrix elements in Eq.
(\ref{A1}), we write the most general form
\begin{eqnarray}\label{A11}
\langle \Lambda\bar p|\bar s \gamma_{\mu}b|B^-\rangle&=& i\bar
u(p_{\Lambda})[a_1\gamma_{\mu}\gamma_5
+a_2p_{\mu}\gamma_5+a_3(p_{\bar p}-p_{\Lambda})\gamma_5]v(p_{\bar
p}),\nonumber\\
\langle \Lambda\bar p|\bar s \gamma_{\mu}\gamma_5b|B^-\rangle&=&
i\bar u(p_{\Lambda})[c_1\gamma_{\mu}
+c_2i\sigma_{\mu\nu}p^\nu+c_3(p_{\bar p}+p_{\Lambda})]v(p_{\bar
p}),
\end{eqnarray}
where $p=p_B-p_{\Lambda}-p_{\bar p}$ and $a_i(c_i)\;(i=1,..., 3)$
are form factors.

To find out the coefficients $a_i(c_i)$ in Eq. (\ref{A11}), we
invoke the work of  Chua,  Hou and  Tsai in Ref.
\cite{transition}. In their analysis,  three form factors $F_A$,
$F_P$ and $F_{V}$ are used to describe
$B\to {\bf B}{\bf\bar B'}$ transitions based on the QCD counting
rules \cite{QCD1}, that require the form factors to behave as
inverse powers of $t=(p_{\bf B}+p_{\bf \bar B'})^2$. The detail
discussions can be  referred to Refs. \cite{transition,QCD}. In
this paper, we shall follow their approach. The representations of
the matrix elements for the $B^-\to p\bar p$ transition are given by
\cite{transition}
\begin{eqnarray}\label{A2}
\langle p \bar p|\bar u(1\pm\gamma_5)b|B^-\rangle&=&i\bar u(p_{p})
[(F_A \not{\!p}\gamma_5\pm F_{V} \not{\! p})+(F_P \gamma_5\pm
F_S)]v(p_{\bar p}),
\end{eqnarray}
with a derived relation $F_S=F_P$. In terms of the approach of
\cite{transition,QCD}, those of the $B^-\to \Lambda\bar p$ transition are given by
\begin{eqnarray}\label{A3}
\langle \Lambda \bar p|\bar s(1\pm\gamma_5)b|B^-\rangle&=&i\bar
u(p_{\Lambda}) [(F_A^{\Lambda \bar p} \not{\! p}\gamma_5\pm
F_{V}^{\Lambda \bar p} \not{\! p}) +(F_P^{\Lambda \bar p}
\gamma_5\pm F_S^{\Lambda \bar p})]v(p_{\bar p}),
\end{eqnarray}
where the form factors related to those of $B^-\to p\bar p$
in Eq. (\ref{A2}) are shown as
\begin{eqnarray}\label{form1}
F^{\Lambda \bar
p}_{A}=\sqrt{\frac{3}{2}}\frac{3}{10}(F_{V}-F_A),\;\; F^{\Lambda
\bar p}_{V}=-\sqrt{\frac{3}{2}}\frac{3}{10}(F_{V}-F_A),\;\;
F^{\Lambda \bar p}_{P(S)}=\sqrt{\frac{3}{2}}\frac{3}{4}F_P.
\end{eqnarray}
The three form factors $F_A$, $F_V$ and $F_P$ can be simply
presented as \cite{transition,QCD}
\begin{eqnarray}\label{para}
F_{A,V}=\frac{C_{A,V}}{t^3},\;\;\;F_{P}=\frac{C_P}{t^4}\,,
\end{eqnarray}
where $C_i\; (i=A,V,P)$ are new parametrized form factors, which are
taking to be real.

From the relation $p^{\mu}\langle \Lambda\bar p|\bar s
\gamma_{\mu}(1-\gamma_5)b|B^-\rangle=m_b\langle \Lambda\bar p|\bar s
(1-\gamma_5)b|B^-\rangle$ in the heavy b quark limit,
 the parameters $a_i(c_i)$ in Eq. (\ref{A11}) are associated with the
scalar and pseudo-scalar matrix elements defined in Eq.
(\ref{A3}). As a result, we get that
\begin{eqnarray}
a_1=m_b F_A^{\Lambda \bar p}\;,\;\;a_3=\frac{m_b F_{P}^{\Lambda
\bar p}}{p\cdot (p_{\bar p}-p_{\Lambda})}\;,\;\;c_1=m_b
F_V^{\Lambda \bar p}\;,\;\;c_3=\frac{m_b F_{P}^{\Lambda \bar
p}}{p\cdot (p_{\bar p}+p_{\Lambda})}.
\end{eqnarray}
The amplitude in Eq. (\ref{A1}) then becomes
\begin{eqnarray}
&&A(B^- \to \Lambda\bar p \gamma)=\frac{G_F}{\sqrt
2}V_{tb}V^*_{ts}\frac{e}{8\pi^2}2c^{eff}_7 \nonumber\\
&&\bigg\{ m_b^3 \varepsilon^{\mu}\; \bar u(p_{\Lambda})
\left[F_A^{\Lambda \bar p}\gamma_{\mu}\gamma_5+F_P^{\Lambda \bar
p}\gamma_5\frac{(p_{\bar p}-p_{\Lambda})_{\mu}}{p\cdot (p_{\bar
p}-p_{\Lambda})}-F_{V}^{\Lambda \bar p}\gamma_{\mu}-F_P^{\Lambda
\bar p}\frac{(p_{\bar p}+p_{\Lambda})_{\mu}}{p\cdot
(p_{\bar p}+p_{\Lambda})}\right]v(p_{\bar p})\nonumber\\
&&-2m_b p_B\cdot\varepsilon \;u(p_{\Lambda}) \left[F_A^{\Lambda
\bar p}\not{\!p}\gamma_5+F_{P}^{\Lambda \bar
p}\gamma_5+F_V^{\Lambda \bar p}\not{\!p}+F_{P}^{\Lambda \bar
p}\right]v(p_{\bar p})\bigg\} , \label{Ampl}
\end{eqnarray}
with three unknown form factors $F^{\Lambda\bar p}_A$, $F^{\Lambda
\bar p}_{V}$ and $F^{\Lambda \bar p}_{P}$. We note that the terms
corresponding to $a_{2}$ disappear due to the fact of
$\varepsilon\cdot p=0$. Even though $c_2$ can only be determined
by experimental data, according to QCD counting rules, $c_2$
needs an additional $1/t$ than $c_1$ to flip the helicity,
so that it
is guaranteed to give a small contribution and can be neglected.

After summing over the photon polarizations and baryon spins,
from Eq. (\ref{Ampl}),
the decay rate  of $\Gamma$  is given by the integration of
\begin{eqnarray}
d \Gamma&=&\frac{1}{(2\pi)^3}\frac{m_b^6}{4
M^3_BE_\gamma^2}|C_t|^2 \left[V|F^{\Lambda\bar
p}_V|^2+A|F^{\Lambda\bar p}_A|^2 +P|F^{\Lambda\bar
p}_P|^2+I_{VP}Re(F^{\Lambda\bar p}_V {F^{\Lambda\bar
p}_P}^*)\right.
\nonumber\\
&&\left. \ \ \ \ \ \ \ \ \ \ \ \ \ \ \ \ \ \ \ \ \ \ \ \ \
+I_{AP}Re(F^{\Lambda\bar p}_A {F^{\Lambda\bar p}_P}^*)\right]
dm^2_{\Lambda \bar p}dm^2_{\bar p\gamma} \label{Rate}
\end{eqnarray}
where
\begin{eqnarray}
m_{\Lambda \bar p} &=& p_{\Lambda}+p_{\bar p}\,,\ \
m_{\bar p\gamma} \;=\; p_{\bar p}+p_{\gamma}\,,\ \
 C_t\,=\,\frac{G_F}{\sqrt 2}V_{tb}V^*_{ts}\frac{e}{8\pi^2}2c^{eff}_7\,,
\nonumber\\
\nonumber\\
 V(A)&=&
 p_{\Lambda}\cdot p(E_{\bar p} E_\gamma-p_{\bar
p}\cdot p)+E_\gamma(E_{\Lambda}p_{\bar p}\cdot p
\pm E_\gamma m_{\Lambda}m_{\bar p}) \,,\nonumber\\
\nonumber\\
P&=& -\frac{E_\gamma(E_\Lambda+E_{\bar p})(m_\Lambda m_{\bar
p}-p_\Lambda\cdot p_{\bar p})}{p_\Lambda\cdot p+p_{\bar p}\cdot p}
+\frac{(m_\Lambda^2+m_{\bar p}^2+2p_\Lambda\cdot p_{\bar
p})(m_\Lambda m_{\bar p}-p_\Lambda\cdot p_{\bar
p})}{2(p_\Lambda\cdot p+p_{\bar p}\cdot p)^2} \nonumber\\&&
+\frac{E_\gamma(E_\Lambda-E_{\bar p})(m_\Lambda m_{\bar
p}-p_\Lambda\cdot p_{\bar p})}{p_\Lambda\cdot p-p_{\bar p}\cdot p}
-\frac{(m_\Lambda^2+m_{\bar p}^2-2p_\Lambda\cdot p_{\bar
p})(m_\Lambda m_{\bar p}+p_\Lambda\cdot p_{\bar
p})}{2(p_\Lambda\cdot p-p_{\bar p}\cdot p)^2} \nonumber\\&&
-p_\Lambda\cdot p_{\bar p}\,,
\nonumber\\
\nonumber\\
I_{VP(AP)}&=& 2E_{\bar p}E_{\gamma}m_{\Lambda}-p_{\bar p}\cdot p
m_{\Lambda}\pm E_{\Lambda}E_{\gamma}(m_{\Lambda}- m_{\bar p})\pm
m_{\bar
p }p_{\Lambda}\cdot p\nonumber\\
&& +\frac{E_\gamma(E_{\bar p}\pm E_{\Lambda})(m_{\Lambda}+m_{\bar
p})p_{\Lambda}\cdot p -E_\gamma^2(m_{\Lambda}-m_{\bar
p})(p_{\Lambda}\cdot p_{\bar p}\pm m_{\Lambda}m_{\bar p}) }
{p_{\Lambda}\cdot p\pm p_{\bar p }\cdot p}\,.
\\
\nonumber
\end{eqnarray}
 It
is important to note that, since the penguin-induced radiative $B$
decays are associated with axial-vector currents  shown in Eq.
(\ref{A1}), we have used \cite{polarization}
\begin{eqnarray}
\sum_{\lambda=1,2}\varepsilon^{*\lambda}_{\mu}\varepsilon_{\nu}^{\lambda}=
-g_{\mu\nu}+\frac{k_{\mu}n_{\nu}+k_{\nu}n_{\mu}}{k\cdot
n}-\frac{k_{\mu}k_{\nu}}{(k\cdot n)^2},
\end{eqnarray}
where $n=(1,0,0,0)$, to sum over the photon polarizations instead
of the direct replacement of
$\sum_{\lambda=1,2}\varepsilon^{*\lambda}_{\mu}
\varepsilon_{\nu}^{\lambda}\rightarrow -g_{\mu\nu}$ which is valid
in  the QED-like theory due to the Ward identity.

For the numerical analysis of the branching ratios, we take the
effective Wilson coefficient $c^{eff}_7=-0.314$ \cite{c7}, the
running quark mass $m_b=4.88$ GeV and CKM matrix elements
$V_{tb}V_{ts}^*=-0.0402$. Even though there are no theoretical
calculations to the unknown $C_A$, $C_V$ and $C_P$. By virtue of
the approach of Ref. \cite{transition}, these form factors are
related to the present experimental data, such as $Br(B^- \to p
\bar p \pi^-)$, $Br(B^0 \to p \bar p K^0)$, $Br(B^- \to p \bar p
K^-)$ \cite{threebody2} and $Br(B^- \to \Lambda \bar \Lambda K^-)$
\cite{threebody3}, characterized by an emitted pseudoscalar meson.
For a reliable $\chi^2$ fitting, we need 2 degrees of freedom (DOF)
by ignoring the $C_P$ term since its contribution is always
associated with
one more $1/t$ over $C_A$ and $C_V$ ones, as seen in Eq.
(\ref{para}). We will take a consistent check in the next
paragraph to this simplification. To illustrate our results, we
fix the color number $N_C=3$ and weak phase
$\gamma=54.8^\circ$. The input experimental data and numerical
values are summarized in Table \ref{parameter}. \footnotesize
\begin{table}[htb]
\begin{center}
\caption{Fits of ($C_A$,$C_V$) in units of $GeV^4$.
 }\label{parameter}
\begin{tabular}{|c|c||c|c|}
\hline Input&experimental data&Fit result&best fit (with 1
$\sigma$ error )\\\hline
$Br(B^- \to p \bar p \pi^-)$ \cite{threebody2}&$3.06\pm0.82$&$C_A$&$-68.3\pm 5.1$\\
$Br(B^0 \to p \bar p K^0)$ \cite{threebody2}&$1.88\pm0.80$&$C_V$&\,\,$35.1\pm 9.0$\\
$Br(B^- \to p \bar p K^-)$ \cite{threebody2}&$5.66\pm0.91$&$\chi^2/DOF$&1.85\\
$Br(B^- \to \Lambda \bar \Lambda K^-)$ \cite{threebody3}&$2.91\pm0.98$&&\\\hline
\end{tabular}
\end{center}
\end{table}
\normalsize

Using the fitted values of $C_A$ and $C_V$, we find $Br(B^- \to
\Lambda \bar p \gamma)=(0.92\pm 0.20)\times 10^{-6}$ which is
larger than its two-body decay partner as expected and it is close
to the result of $1.2\times 10^{-6}$ in the pole model \cite{HY}.
However, our predicted value on $B^- \to \Lambda \bar p \gamma$ is
smaller than $(2.16^{+0.58}_{-0.53}\pm 0.20)\times 10^{-6}$
\cite{radiative} measured by BELLE. If we put this new observed
value into our fitting, we can further include $C_P$ ignored
previously. The fitted values  are $C_A=-73.3\pm 9.1\ GeV^4$,
$C_V=43.7\pm 12.1\ GeV^4$ and $C_P=134.3\pm 327.0\ GeV^7$
 with $\chi^2/DOF=3.65$
which is about two times bigger than previous one. Clearly, it
presents an inferior fitting with small $C_{A,V}$ changes. When
putting back these three fitted values to $Br(B^- \to \Lambda \bar
p \gamma)$ for a consistency check, we get $(1.16\pm 0.31)\times
10^{-6}$ regardless of inputting larger experimental value, which
explains the large value of $\chi^2/DOF$. The insensitivity of
$C_P$ on the decay branching ratio justifies our early
simplification of ignoring its contribution beside the $1/t$
argument.

\begin{figure}[t!]
\centering
\includegraphics[width=3.5in]{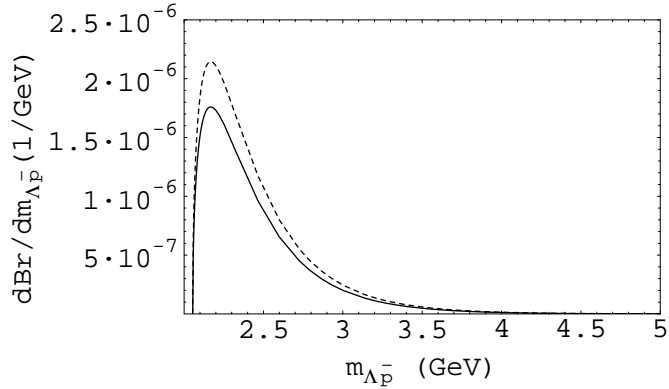}
\caption{\label{br12} $dBr(B^- \to \Lambda\bar
p\gamma)$/$dm_{\Lambda \bar p}$ vs. $m_{\Lambda \bar p}$. The
solid line stands for the input values of $(C_A,C_V)=(-68.3,35.1)$
while the dash line stands for those of
$(C_A,C_V,C_P)=(-73.3,43.7,134.3)$. }
\end{figure}

In Ref. \cite{threshold}, it was suggested that the reduced energy
release can make the branching ratios of three-body decays as
significant as their counterparts of two-body modes or even
larger, and one of the signatures would be baryon pair threshold
effect \cite{threshold,transition}. In Fig.  \ref{br12}, from Eq.
(\ref{Rate}) we show the differential branching ratio of $dBr(B^-
\to \Lambda\bar p\gamma)/dm_{\Lambda \bar p}$ vs. $m_{\Lambda\bar
p}$ representing the threshold enhancement around the invariant mass $m_{\Lambda\bar{p}}=2.05$ GeV, which is consistent with Fig. 2 in Ref.
\cite{radiative} of the BELLE result. Around the threshold, the baryon
pair contains  half of the $B$ meson energy while the phone
emitting back to back to the baryon pair with another half
of energy which explains the peak at $E_\gamma\sim 2\ GeV$ in Fig. 3 of Ref. \cite{radiative}.
Such mechanism is similar to the two-body decays so that
factorization method works \cite{threshold} even in the three-body
decays.

To discuss other radiative baryonic $B^-$ decays, we give form
factors by relating them to $F_{V,A,P}$ in the $B^- \to p \bar p$
transition similar to the case of $B^-\to\Lambda\bar{p}\gamma$
as follows:
\begin{eqnarray}
&&B^- \to \Sigma^0 \bar p \gamma:\nonumber\\
&&F^{\Sigma^0 \bar p}_{V}=-\frac{11 F_{V}}{10\sqrt 2}-\frac{ 9
F_A}{10\sqrt 2},\; F^{\Sigma^0 \bar p}_{A}= -\frac{9
F_{V}}{10\sqrt 2}-\frac{11F_A}{10\sqrt 2},\; F^{\Sigma^0 \bar
p}_{P}=\frac{F_{P}}{3\sqrt 2},\;\nonumber
\end{eqnarray}
\begin{eqnarray}
&&B^- \to \Sigma^-\bar n \gamma :\nonumber\\
&&F^{\Sigma^-\bar n}_{V}=-\frac{11F_{V}}{10}-\frac{9F_A}{10},\;
F^{\Sigma^-\bar n}_{A}=-\frac{9F_{V}}{10}-\frac{9F_A}{11},\;
F^{\Sigma^-\bar n}_{P}=\frac{F_P}{4},\;\;\;\;\;\;\;\;\nonumber
\end{eqnarray}
\begin{eqnarray}
&&B^- \to \Xi^- \bar \Lambda  \gamma:\nonumber\\
&&F^{\Xi^- \bar \Lambda }_{V}=-\frac{21F_{V}}{10\sqrt
6}-\frac{9F_A}{10\sqrt 6},\; F^{\Xi^- \bar \Lambda
}_{A}=-\frac{9F_{V}}{10\sqrt 6}-\frac{21F_A}{10\sqrt 6},\;
F^{\Xi^- \bar \Lambda }_{P}=\frac{F_P}{4},\nonumber
\end{eqnarray}
\begin{eqnarray}
&&B^- \to \Xi^0 \bar \Sigma^-   \gamma :\nonumber\\
&&F^{\Xi^0 \bar \Sigma^-
}_{V}=-\frac{F_{V}}{10}-\frac{9F_A}{10},\; F^{\Xi^0 \bar \Sigma^-
}_{A}=-\frac{9F_{V}}{10}-\frac{F_A}{10},\; F^{\Xi^0 \bar \Sigma^-
}_{P}=\frac{5F_P}{4},\;\;\;\;\;\;\;\;\nonumber
\end{eqnarray}
\begin{eqnarray}
\label{formtotal}
&&B^- \to \Xi^- \bar \Sigma^0  \gamma:\nonumber\\
&&F^{\Xi^- \bar \Sigma^0
}_{V}=-\frac{F_{V}}{10\sqrt{2}}-\frac{9F_A}{10\sqrt{2}},\;
F^{\Xi^- \bar \Sigma^0
}_{A}=-\frac{9F_{V}}{10\sqrt{2}}-\frac{F_A}{10\sqrt{2}}, F^{\Xi^-
\bar \Sigma^0 }_{P}=\frac{5F_P}{4\sqrt 2}.
\end{eqnarray}
To calculate the branching ratio of $B\to {\bf B}{\bf \bar
B'}\gamma$, we can use the formula in Eq. (\ref{Rate}) by
replacing $\Lambda$ and $\bar p$ by ${\bf B}$ and ${\bf \bar B'}$,
respectively. The two sets of predicted values for $B\to {\bf
B}{\bf \bar B'}\gamma$ with and without $C_P$ are shown in Table
\ref{brtotal}, respectively. As a comparison, we also list the
work of the pole model approach by Cheng and Yang \cite{HY} in the
table.
\begin{table}[htb]
\footnotesize
\begin{center}
\caption{Decay branching ratios}\label{brtotal}
\begin{tabular}{|l|c|c|c|}
\hline
&\multicolumn{2}{c}{Fits}\vline&\\
\cline{2-3} 
{\bf Branching Ratios}&$(C_A,C_V)=$                  &$(C_A,C_V,C_P)=$                            &
{Pole model\cite{HY}}\\
                                   &$(-68.3\pm 5.1 ,35.0\pm 9.0 )$&$(-73.3\pm 9.1,43.7\pm 12.1,134.3\pm 327.0)$&\\\hline
$Br(B^- \to \Lambda\bar p \gamma)$
&$(0.92\pm 0.20) \times 10^{-6}$&[$(1.16\pm 0.31)\times 10^{-6}$]&$1.2\times 10^{-6}$\\
$Br(B^- \to \Sigma^0 \bar p \gamma)$
&$(1.7\pm 1.5)\times 10^{-7}$&$(1.2\pm 1.2)\times 10^{-7}$&$2.9\times 10^{-9}$\\
$Br(B^- \to \Sigma^-\bar n \gamma)$
&$(3.4\pm 2.8)\times 10^{-7}$&$(2.5\pm 2.4)\times 10^{-7}$&$5.7\times 10^{-9}$\\
$Br(B^- \to \Xi^- \bar \Lambda \gamma)$
&$(0.48\pm 0.50)\times 10^{-7}$&$(0.61\pm 0.60)\times 10^{-7}$&$2.4\times 10^{-7}$\\
$Br(B^- \to \Xi^0 \bar \Sigma^- \gamma)$
&$(3.3\pm 0.7)\times 10^{-7}$&$(3.7\pm 0.9)\times 10^{-7}$&$1.2\times 10^{-6}$\\
$Br(B^- \to \Xi^- \bar \Sigma^0 \gamma)$
&$(1.5\pm 0.6)\times 10^{-7}$&$(1.8\pm 0.6)\times 10^{-7}$&$6.0\times 10^{-7}$\\\hline
\end{tabular}
\end{center}
\end{table}
We note that, in Table \ref{brtotal}, the value in the bracket of
the third column for $Br(B^- \to \Lambda \bar p \gamma)$ is not a
prediction but a consistency comparison with the puting-back form
factors, since we have used the observed value of $Br(B^- \to
\Lambda \bar p \gamma)$ from BELLE. We found that, except for
$Br(B^- \to \Lambda \bar p \gamma)$, all predicted values are
  $O(10^{-7})$. In terms of inverse sign between $C_A$ and
$C_V$, there are constructive effects for $F^{\Lambda \bar p}_A$
and $F^{\Lambda \bar p}_{V}$, which are proportional to
($F_{V}-F_A$) as shown in Eq. (\ref{form1}), whereas destructive
effects make other $F^{{\bf B} {\bf\bar B'}}_A$ and $F^{{\bf B}
{\bf\bar B'}}_V$ in Eq. (\ref{formtotal}) small. Consequently, all
modes for $B^-$ radiative baryonic decays are suppressed except
for $Br(B^- \to \Lambda \bar p \gamma)$. We remark that such
suppressions exist only in the SM-like theories. Thus, these
radiative baryonic decays are useful modes for testing the new
physics.

As seen in Table \ref{brtotal}, both our results and those of the
pole model satisfy the relations of $Br(B^- \to \Sigma^-\bar
n\gamma) \simeq 2 Br(B^- \to \Sigma^0\bar p\gamma)$ and
$Br(B^-\to\Xi^0 \bar \Sigma^- \gamma)\simeq 2 Br(B^- \to \Xi^-
\bar \Sigma^0\gamma)$ because of the SU(3) symmetry. In the pole
model, the decay branching ratios of $B^- \to \Lambda \bar p
\gamma$ and $B^- \to \Xi^0 \bar \Sigma^- \gamma$  are found to be
large, around $1.2 \times 10^{-6}$, since ther are intermediated
through $\Lambda_b$ and $\Xi_b$, which correspond to large
coupling constants $g_{\Lambda_b \to B^-p}$ and $g_{\Xi^0_b \to
B^- \Sigma^+}$, respectively.
 However, in our work, the branching ratio of
$B^- \to\Lambda\bar p \gamma$ is about three times larger than that
of $B^- \to \Xi^0 \bar \Sigma^- \gamma$, which is $O(10^{-7})$. Regardless
of these differences, both two methods are within the experimental
data allowed ranges, such as those of
\begin{eqnarray}
&[Br(B^- \to \Lambda \bar p \gamma)+0.3
Br(B^- \to \Sigma^0 \bar p\gamma)]_{E_{\gamma}>2.0\text{ GeV}}<3.3\times 10^{-6}\,,\nonumber\\
&[Br(B^- \to \Sigma^0 \bar p\gamma)+0.4 Br(B^- \to \Lambda \bar p
\gamma)]_{E_{\gamma}>2.0\text{ GeV}}<6.4\times 10^{-6}\,,\nonumber
\end{eqnarray}
from CLEO \cite{cleo} and $Br(B^- \to \Sigma^0 \bar p\gamma)<3.3\times
10^{-6}$  from BELLE \cite{radiative}.

Finally, we relate the $\bar B^0$ decays with the
corresponding $B^-$ modes in terms of QCD counting rules even though
there are no experimental data on radiative baryonic $\bar B^0$ decays. When
neglecting the mass and life time differences, we obtain
\begin{eqnarray}\label{relation}
Br(B^- \to \Lambda \bar p \gamma)=Br(\bar B^0 \to \Lambda \bar n\gamma),&
Br(B^- \to \Sigma^0 \bar p \gamma)=Br(\bar B^0 \to \Sigma^0 \bar n \gamma),\nonumber\\
Br(B^- \to \Sigma^-\bar n \gamma)=Br(\bar B^0 \to \Sigma^+\bar p\gamma),&
Br(B^- \to \Xi^- \bar \Lambda \gamma)=Br(\bar B^0 \to \Xi^0 \bar \Lambda \gamma),\nonumber\\
Br(B^- \to \Xi^- \bar \Sigma^0 \gamma)=Br(\bar B^0 \to \Xi^0 \bar \Sigma^0 \gamma),&
Br(B^- \to \Xi^0 \bar \Sigma^- \gamma)=Br(\bar B^0 \to \Xi^- \bar \Sigma^+\gamma)\,,
\end{eqnarray}
which are also guaranteed by the $SU(3)$ symmetry.
From Eq. (\ref{relation}), we see that $Br(\bar B^0 \to \Lambda
\bar n\gamma)$ can  be as large as
$Br(B^- \to \Lambda \bar p \gamma)$.

In sum, we have shown that the SD contributions to the radiative
baryonic decays of $B \to {\bf B}{\bf \bar B'}\gamma$ in the
SM are  associated with the form factors of $F_A$,
$F_{V}$ and $F_P$ in the matrix elements of
 the $B^- \to p \bar
p$ transition. Most of the predicted values for $Br(B\to {\bf B}{\bf\bar
B'}\gamma)$ are spanning in the order of $10^{-7}$, which are
larger than the estimated values of $O(10^{-9})$ due to the IB
effects of their two-body counterparts. In particular, we have
found that $Br(B^- \to \Lambda \bar p\gamma)$
is  $(1.16\pm0.31)\times 10^{-6}$ and $(0.92\pm0.20)\times 10^{-6}$
with and without $C_P$, respectively,
which are
consistent with the pole model prediction \cite{HY} but smaller than the experimental data from BELLE \cite{radiative}.
More precise measurements are clearly needed.

\section*{Acknowledgements}

We would like to thank C. K. Chua and S. Y. Tsai for
illuminating and useful discussions.
This work was supported in part by
 the National Science Council of the Republic of China under
 Contract No. NSC-92-2112-M-007-025.

\end{document}